\documentclass[useAMS,usenatbib]{mn2e}

\usepackage{natbib}
\usepackage{amsmath,amsfonts,amssymb}
\usepackage[dvips]{graphicx}
\usepackage{longtable,multirow}
\usepackage{float}
\usepackage{caption}
\usepackage{subcaption}
\usepackage{txfonts}
\usepackage{color,psfrag}
\usepackage{afterpage}

\title[AMI Galactic Plane Survey -- Data Analysis]
  {AMI Galactic Plane Survey at 16~GHz: I -- Observing, mapping and source extraction\thanks{We request that any reference to this paper cites `AMI Consortium: Perrott et al. 2012'}}
   
    \author[Perrott et~al.]{AMI Consortium: Yvette~C.~Perrott$^{1}\thanks{Corresponding author: email -- ycp21@mrao.cam.ac.uk}$, Anna~M.~M.~Scaife$^{2}$, David~A.~Green$^{1}$, \newauthor
    Matthew~L.~Davies$^{1}$, Thomas~M.~O.~Franzen$^{3}$, Keith~J.~B.~Grainge$^{1,4}$, \newauthor
    Michael~P.~Hobson$^{1}$, Natasha~Hurley-Walker$^{5}$, Anthony~N.~Lasenby$^{1,4}$, \newauthor
    Malak~Olamaie$^{1}$, Guy~G.~Pooley$^{1}$, Carmen~Rodr\'{i}guez-Gonz\'{a}lvez$^{6}$, \newauthor
    Clare~Rumsey$^{1}$, Richard~D.~E.~Saunders$^{1,4}$, Michel~P.~Schammel$^{1}$, \newauthor
    Paul~F.~Scott$^{1}$, Timothy~W.~Shimwell$^{3}$,  David~J.~Titterington$^{1}$, Elizabeth~M.~Waldram$^{1}$\\ 
 $^1$ Astrophysics Group, Cavendish Laboratory,
      19 J.~J.~Thomson Avenue, Cambridge CB3 0HE \\
 $^2$ School of Physics and Astronomy, University of Southampton, Highfield, Southampton, SO17 1BJ \\
 $^3$ CSIRO Astronomy \& Space Science, Australia Telescope National Facility, PO Box 76, Epping, NSW 1710, Australia \\
 $^4$ Kavli Institute for Cosmology Cambridge,
      Madingley Road, Cambridge CB3 0HA\\
 $^5$ International Centre for Radio Astronomy Research, Curtin Institute of Radio Astronomy, 1 Turner Avenue, Technology Park, Bentley, WA 6845, Australia \\
 $^6$ Spitzer Science Center, MS 220-6, California Institute of Technology, Pasadena, CA 91125 USA
 \\
}

\date{Accepted 2012 December 10. Received 2012 December 10}
\pagerange{\pageref{firstpage}--\pageref{lastpage}}
\pubyear{2012}

\begin{document}
\maketitle
\label{firstpage}

\begin{abstract}

\noindent
The AMI Galactic Plane Survey (AMIGPS) is a large area survey of the outer
Galactic plane to provide arcminute resolution images at milli-Jansky
sensitivity in the centimetre-wave band. Here we present the first data release of the survey, consisting of 868\,deg$^2$ of the
Galactic plane, covering the area $76^{\circ} \lessapprox \ell \lessapprox 170^{\circ}$
between latitudes of $|b| \lessapprox 5^{\circ}$, at a central frequency of 15.75\,GHz
(1.9\,cm).  We describe in detail the drift scan
observations which have been used to construct the maps, including the
techniques used for observing, mapping and source extraction, and summarise the
properties of the finalized datasets.  These observations constitute the most sensitive Galactic plane survey of large
extent at centimetre-wave frequencies greater than 1.4\,GHz.
\end{abstract}


%%%%%%%%%%%%%%%%%%%%%%%%%%%%%%%%%%%%%%%%%%%%%%%%%%%%%%%%%%%%%%%%%%%
\section{Introduction}\label{Introduction}
%%%%%%%%%%%%%%%%%%%%%%%%%%%%%%%%%%%%%%%%%%%%%%%%%%%%%%%%%%%%%%%%%%%

Large-area radio surveys contribute to our understanding of the Universe in
numerous and diverse ways. Discoveries from these surveys
have become key ingredients of modern astrophysics: pulsars, radio galaxies and
quasars and more (see e.g. \citealt{1998ASSL..226....3L}).  For studies of our Galaxy, radio surveys are particularly beneficial as the longer
wavelength radio emission does not suffer from the same extinction and opacity
effects as optical and infra-red surveys and the dense regions of dust and gas
which dominate the low-latitude Galactic plane become largely transparent,
allowing us to study sources in these regions.  However, the bulk of Galactic radio surveys are at frequencies at or below 1.4\,GHz and as such are necessarily biased against objects whose spectra rise with frequency, such as dense star-forming regions.  Two examples of the need for higher-frequency, centimetre-wave Galactic surveying are as follows.

The first is the hypercompact {\sc Hii} ({\sc HCHii}) region. Thought to indicate the earliest visible stage of massive star formation, these objects are two orders of magnitude more dense than the better known ultracompact ({\sc UCHii}) region and have steeply rising spectra. {\sc
HCHii} regions were discovered serendipitously in observations of {\sc UCHii},
having been missed previously in their entirety by Galactic plane surveys
concentrated at $\nu<5$\,GHz. The turnover frequency between the optically
thick and thin regimes for thermal bremsstrahlung is a linear function of
emission measure (e.g. \citealt{1967ApJ...147..471M}) causing such low frequency surveys
(e.g. $\nu\leq 5$\,GHz) to preferentially select against dense plasmas ($n_{\rm
e} \leq 10^{11}$\,m$^{-3}$). Such plasmas are not limited to {\sc HCHii} regions
but also include a variety of other Galactic objects such as massive stellar
winds, ionised jets from young stellar objects (e.g. \citealt{1995RMxAC...1...67A}) and young planetary nebulae (e.g. \citealt{2009MNRAS.397.1386B}).

The second is the anomalous microwave emission (AME), now being identified in an increasing
number of Galactic objects, that was missed in low frequency Galactic surveys. First identified by CMB experiments \citep{1997ApJ...486L..23L} as a large scale foreground contaminant this form of emission has since been demonstrated to exist in more compact objects such as dark (e.g. \citealt{2006ApJ...639..951C}; \citealt{2009MNRAS.394L..46A}; \citealt{2010MNRAS.403L..46S}) and molecular clouds (\citealt{2005ApJ...624L..89W}; \citealt{2011MNRAS.418.1889T}). Although multiple mechanisms have been proposed to explain AME, dipole emission from rapidly rotating very small dust grains (\citealt{1998ApJ...494L..19D}, \citealt{1998ApJ...508..157D}) is generally considered to be most likely. Such spinning dust
emission has a peaked SED with a maximum in the frequency range
10 -- 50\,GHz depending on grain size distributions. 

A current lack of surveys in
this frequency range means that our knowledge of the overall properties of
objects which exhibit emission from spinning dust, objects which are
characterized by dense plasmas, and indeed the global distribution of rising-spectrum emission in the Galaxy, is extremely poor. Those surveys which are
available, such as the 9C Ryle Telescope survey (15\,GHz; \citealt{2003MNRAS.342..915W}),
the GPA survey (14.35\,GHz; \citealt{2000AJ....119.2801L}) and the AT20G survey
(20\,GHz; \citealt{2010MNRAS.402.2403M}) have provided us with tantalising insights into
the high frequency Galactic plane, but there is a continuing need for higher
sensitivity, resolution and sky area coverage at these frequencies.

The interferometric Arcminute Microkelvin Imager (AMI) Galactic Plane Survey (AMIGPS) provides
the most sensitive centimetre-wave Galactic plane survey of large extent at $\nu > 1.4$\,GHz.  AMIGPS is a drift-scan survey of the northern Galactic plane at $\approx16$\,GHz, covering (in the first data release) the region $76^{\circ} \lessapprox \ell \lessapprox 170^{\circ}$ and $|b|\lessapprox 5^{\circ}$.  The AMI Small Array (SA) has been used for the survey since its relatively large field of view ($\approx400$\,arcmin$^{2}$) makes covering large areas feasible, and its short baselines mean that extended objects, very common in the Galaxy, are at least partially observable.  The resolution of the survey is $\approx3$\,arcmin and the noise level is $\approx3$\,mJy\,beam$^{-1}$ away from bright sources.

This paper focuses on the techniques employed for observing (Section~\ref{Observations}), mapping (Section~\ref{Data reduction and mapping})
and source extraction (Section~\ref{Source extraction}) in the AMIGPS.  The positional and flux density calibration accuracy of the survey are also tested in Section~\ref{Calibration accuracy checks}, and in Section~\ref{Data products} the maps and catalogue are described.  In a following paper, hereafter Paper~II, the first results from the
survey, including the follow-up of rising-spectrum objects in order to detect \textsc{U/HCHii} regions, will be presented.

%%%%%%%%%%%%%%%%%%%%%%%%%%%%%%%%%%%%%%%%%%%%%%%%%%%%%%%%%%%%%%%%%%%
\section{The Arcminute Microkelvin Imager Small Array}\label{The Arcminute Microkelvin Imager Small Array}
%%%%%%%%%%%%%%%%%%%%%%%%%%%%%%%%%%%%%%%%%%%%%%%%%%%%%%%%%%%%%%%%%%%

The AMI (\citealt{zwart2008}) was designed as a Sunyaev--Zel'dovich (SZ) effect instrument, and consists of two separate instruments: the Small Array (SA), optimised for observing extended SZ decrements on arcminute scales, and
the Large Array (LA) with higher resolution (30\,arcsec), used for characterising and subtracting point source foregrounds from SA data.  The AMIGPS was carried out solely with the SA, although some follow-up observations were also made using the LA; these latter will be used as calibration accuracy checks in Section~\ref{Calibration accuracy checks} and will be described fully in Paper~II.

The SA is an interferometer array comprising ten 3.7-metre-diameter,
equatorially-mounted dishes, with a range of baselines of 5 -- 20\,m.
It operates over frequencies 13.9
-- 18.2\,GHz with the passband divided into six channels of 0.72-GHz
bandwidth.  It has a primary beam at the central frequency of 15.75\,GHz of 
$\approx20$\,arcmin full width at half-maximum (FWHM) and a typical
synthesised beam FWHM of $\approx 3$\,arcmin (this varies depending on the
precise \textit{uv}-coverage of any observation).  The telescope
measures a single, linear polarisation (Stokes $I + Q$)
and has a flux-density sensitivity of $\approx 30$\,mJy\,s$^{-1/2}$.  It is sensitive to angular scales of up to $\approx 10$\,arcmin (depending on the \textit{uv}-coverage).

%%%%%%%%%%%%%%%%%%%%%%%%%%%%%%%%%%%%%%%%%%%%%%%%%%%%%%%%%%%%%%%%%%%
\section{Observations}\label{Observations}
%%%%%%%%%%%%%%%%%%%%%%%%%%%%%%%%%%%%%%%%%%%%%%%%%%%%%%%%%%%%%%%%%%%

The AMIGPS is observed in drift-scan mode, in which the SA is pointed at a fixed azimuth and elevation and observes a narrow strip as the sky drifts past.  In practice, to enable re-observation of strips as necessary, the telescope is driven very slowly to maintain a constant J2000 declination.  In order to perform phase calibration, bright nearby point sources selected from the Very Long Baseline Array Calibrator Survey (VCS, \citealt{2002ApJS..141...13B}) were observed for 400\,seconds at 30-minute intervals during each scan.  Strips are observed at a separation of 12\,arcmin in $\delta$, corresponding to the 35\% point of the power primary beam, i.e. at distance $x$ from the centre where $\exp(-x^{2}/(2\sigma^{2}))=0.35$, assuming the beam is Gaussian with width $\sigma$.  This produces a very even noise level across the combined map, with a variation of $\approx3$\% between the centre of a declination strip and the point halfway between declination strips.  The noise level in the survey is $\approx3$\,mJy\,beam$^{-1}$ away from bright sources and is as low as $\approx1$\,mJy\,beam$^{-1}$ at some points.

\begin{figure*}
  \begin{center}
    % This file is generated by the MATLAB m-file laprint.m. It can be included
% into LaTeX documents using the packages graphicx, color and psfrag.
% It is accompanied by a postscript file. A sample LaTeX file is:
%    \documentclass{article}\usepackage{graphicx,color,psfrag}
%    \begin{document}\input{gal_coverage}\end{document}
% See http://www.mathworks.de/matlabcentral/fileexchange/loadFile.do?objectId=4638
% for recent versions of laprint.m.
%
% created by:           LaPrint version 3.16 (13.9.2004)
% created on:           06-Aug-2012 17:31:45
% eps bounding box:     15 cm x 11.25 cm
% comment:              
%
\begin{psfrags}%
\psfragscanon%
%
% text strings:
\psfrag{s05}[t][t]{\color[rgb]{0,0,0}\setlength{\tabcolsep}{0pt}\begin{tabular}{c}Galactic longitude (degrees)\end{tabular}}%
\psfrag{s06}[b][b]{\color[rgb]{0,0,0}\setlength{\tabcolsep}{0pt}\begin{tabular}{c}Galactic latitude (degrees)\end{tabular}}%
\psfrag{s10}[][]{\color[rgb]{0,0,0}\setlength{\tabcolsep}{0pt}\begin{tabular}{c} \end{tabular}}%
\psfrag{s11}[][]{\color[rgb]{0,0,0}\setlength{\tabcolsep}{0pt}\begin{tabular}{c} \end{tabular}}%
%
% xticklabels:
\psfrag{x01}[t][t]{0}%
\psfrag{x02}[t][t]{0.1}%
\psfrag{x03}[t][t]{0.2}%
\psfrag{x04}[t][t]{0.3}%
\psfrag{x05}[t][t]{0.4}%
\psfrag{x06}[t][t]{0.5}%
\psfrag{x07}[t][t]{0.6}%
\psfrag{x08}[t][t]{0.7}%
\psfrag{x09}[t][t]{0.8}%
\psfrag{x10}[t][t]{0.9}%
\psfrag{x11}[t][t]{1}%
\psfrag{x12}[t][t]{310}%
\psfrag{x13}[t][t]{  0}%
\psfrag{x14}[t][t]{ 50}%
\psfrag{x15}[t][t]{100}%
\psfrag{x16}[t][t]{150}%
\psfrag{x17}[t][t]{200}%
\psfrag{x18}[t][t]{250}%
%
% yticklabels:
\psfrag{v01}[r][r]{0}%
\psfrag{v02}[r][r]{0.1}%
\psfrag{v03}[r][r]{0.2}%
\psfrag{v04}[r][r]{0.3}%
\psfrag{v05}[r][r]{0.4}%
\psfrag{v06}[r][r]{0.5}%
\psfrag{v07}[r][r]{0.6}%
\psfrag{v08}[r][r]{0.7}%
\psfrag{v09}[r][r]{0.8}%
\psfrag{v10}[r][r]{0.9}%
\psfrag{v11}[r][r]{1}%
\psfrag{v12}[r][r]{-10}%
\psfrag{v13}[r][r]{-5}%
\psfrag{v14}[r][r]{0}%
\psfrag{v15}[r][r]{5}%
\psfrag{v16}[r][r]{10}%
\psfrag{v17}[r][r]{15}%
\psfrag{v18}[r][r]{20}%
%
% Figure:
\resizebox{0.9\linewidth}{!}{\includegraphics{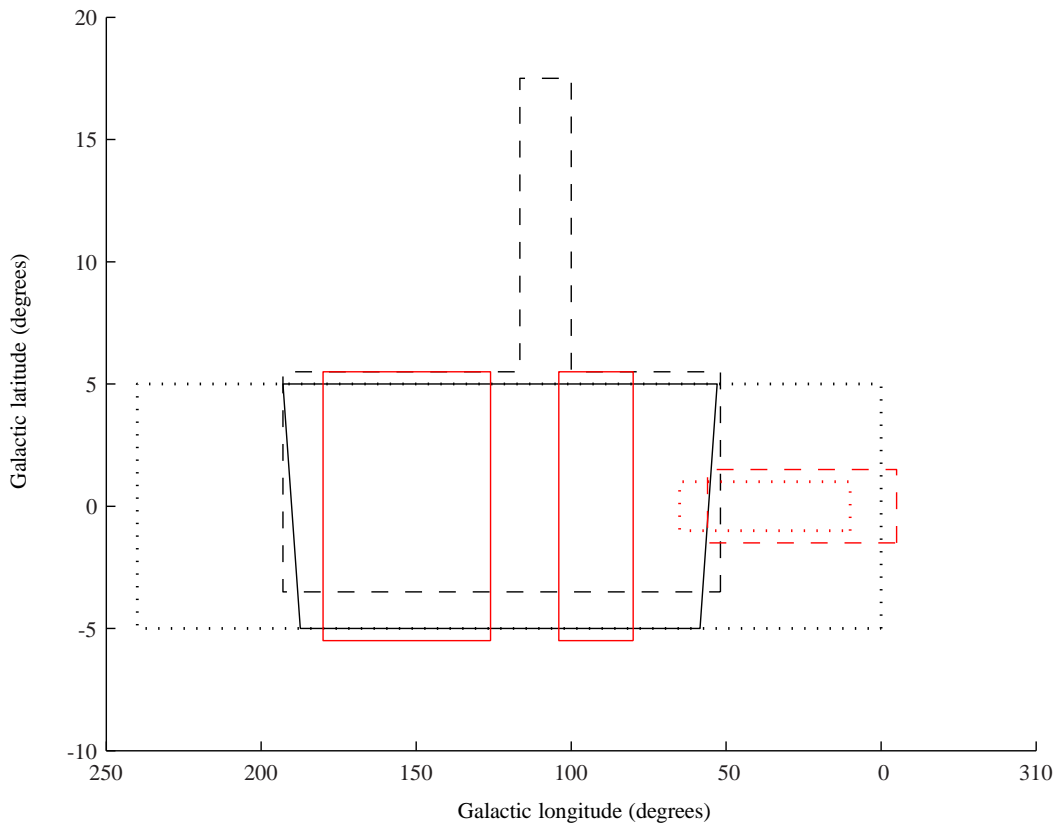}}%
\end{psfrags}%
%
% End gal_coverage.tex

    \caption{Coverage of the full AMIGPS compared to other northern Galactic plane surveys of similar area and/or resolution and noise level.  The AMIGPS boundaries are shown as a solid black line, CGPS (408, 1420\,MHz) as a dashed black line, Effelsberg (1.4, 2.7\,GHz) as a dotted black line, 7C(G) (151\,MHz) as a solid red line, Nobeyama (10\,GHz) as a dashed red line, and CORNISH (5\,GHz) as a dotted red line.}
    \label{Fi:coverage}
  \end{center}
\end{figure*}

This data release consists of observations above $\delta = 40^{\circ}$ and between $b \approx \pm5^{\circ}$; a later data release will extend the coverage to $\delta \geq 20^{\circ}$, corresponding to $53^{\circ} \lessapprox \ell \lessapprox 76^{\circ}$ and $170^{\circ} \lessapprox \ell \lessapprox 193^{\circ}$\,.  The coverages of some other, currently available Galactic plane surveys along with their resolutions and noise levels are shown in Table~\ref{tab:gal_surveys}, and some of these are illustrated in comparison to the (full) AMIGPS in Fig.~\ref{Fi:coverage}.  The AMIGPS is the first survey at this frequency to achieve similar coverage area, resolution and noise level to lower frequency surveys such as the Canadian Galactic Plane Survey (CGPS; \citealt{2003AJ....125.3145T}); earlier surveys have either been wide and shallow with lower resolution, e.g. the GPA, \emph{Wilkinson Microwave Anisotropy Probe} (\emph{WMAP}; ~\citealt{2003ApJ...583....1B}), and \emph{Planck}~\citep{2011A&A...536A...7P}, or narrower, with comparable resolution but still more shallow than the AMIGPS, e.g. Nobeyama at 10\,GHz \citep{1987PASJ...39..709H}.

\begin{table}
\centering
\caption{Coverage, resolution and noise levels of selected Galactic plane surveys.  The noises level marked with (*) are actually detection limits.} \label{tab:gal_surveys}
\begin{tabular}{ccccc}\hline
Telescope/ & Frequency & Coverage & Resolution & Noise level  \\
Survey name & (GHz) & (deg$^2$) & (arcmin) & (mJy\,beam$^{-1}$) \\ \hline
7C(G)$^{a}$ & 0.151 & 1700 & $1.17\,\mathrm{cosec}(\delta)$ & 40 \\[3pt]
AT20G$^{b}$ & 20 & 20086 & 1.7 & 10 \\[3pt]
\multirow{2}{*}{CGPS$^{c}$} & 1.42 & \multirow{2}{*}{1500} & $1\,\mathrm{cosec}(\delta)$ & 0.23 \\
 & 0.408 & & $3.4\,\mathrm{cosec}(\delta)$ & 3 \\[3pt]
CORNISH$^{d}$ & 5 & 110 & 0.017 & 0.4 \\[3pt]
\multirow{3}{*}{Effelsberg} & 4.875$^{e}$ & 125 & 2.6 & 120(*) \\
 & 1.4$^{f,g}$ & \multirow{2}{*}{2400} & 9.3 & 80 \\
 & 2.7$^{h,i}$ & & 4.3 & 50 \\[3pt]
\multirow{2}{*}{GPA$^{j}$} & 8.35 & \multirow{2}{*}{2700} & 10 & 230 \\
 & 14.35 & & 7 & 800 \\[3pt]
MAGPIS$^{k}$ & 1.42 & 43.2 & 0.083 & 0.2 \\[3pt]
Nobeyama$^{l}$ & 10 & 183 & 3 & 33 \\[3pt]
\multirow{3}{*}{RATAN$^{m}$} & 0.96 & \multirow{3}{*}{400} & $4\times 75$ & 60 \\
 & 3.9 & & $1\times 39$ & 10 \\
 & 11.2 & & $0.35\times 14$ & 100 \\[3pt]
\emph{Planck} LFI$^{n}$ & 30 -- 70 & all-sky & 13--33 & 480--585(*) \\
\emph{Planck} HFI$^{n}$ & 100 -- 857 & all-sky & 4--10 & 183--655(*) \\[3pt]
Stockert$^{o}$ & 2.72 & 10200 & 18 & 140 \\[3pt]
VGPS$^{p}$ & 1.42 & $< 200$ & 1 & 2 \\ [3pt]
\multirow{2}{*}{VLA} & 5$^{q}$ & 40 & 0.07 & 2.5--10 \\
 & 1.42$^{r}$ & 224 & 0.07 & 10 \\[3pt]
VSA$^{s}$ & 30 & 152 & 13 & 90 \\[3pt]
WMAP$^{t,u}$ & 23--94 & all-sky & 13--53 & 200--400(*) \\ \hline
\noalign{$^{a}$ \citet{1998MNRAS.294..607V}; $^{b}$ \citet{2010MNRAS.402.2403M}; $^{c}$ \citet{2003AJ....125.3145T};}
\noalign{$^{d}$ \citet{2008ASPC..387..389P}; $^{e}$ \citet{1979A&AS...35...23A};}
\noalign{$^{f}$ \citet{1990A&AS...83..539R}; $^{g}$ \citet{1997A&AS..126..413R};}
\noalign{$^{h}$ \citet{1984A&AS...58..197R}; $^{i}$ \citet{1990A&AS...85..633R}; $^{j}$ \citet{2000AJ....119.2801L};}
\noalign{$^{k}$ \citet{2006AJ....131.2525H}; $^{l}$ \citet{1987PASJ...39..709H}; $^{m}$ \citet{1998IAUS..179..103T};}
\noalign{$^{n}$ \citet{2011A&A...536A...7P}; $^{o}$ \citet{1987MitAG..70..419R};}
\noalign{$^{p}$ \citet{2006AJ....132.1158S}; $^{q}$ \citet{1994ApJS...91..347B};}
\noalign{$^{r}$ \citet{1990ApJS...74..181Z}; $^{s}$ \citet{2010MNRAS.406.1629T};}
\noalign{$^{t}$ \citet{2003ApJ...583....1B}; $^{u}$ \citet{2011ApJS..192...15G}}

\end{tabular}
\end{table}

The observations for the first data release were performed between 22 Jun 2010 and 4 Nov 2011.  Approximately two thirds of the strips were observed multiple times in order to improve the noise level, resulting in a total observing time of $\approx1200$\,hours.

%%%%%%%%%%%%%%%%%%%%%%%%%%%%%%%%%%%%%%%%%%%%%%%%%%%%%%%%%%%%%%%%%%%
\section{Data reduction and mapping}\label{Data reduction and mapping}
%%%%%%%%%%%%%%%%%%%%%%%%%%%%%%%%%%%%%%%%%%%%%%%%%%%%%%%%%%%%%%%%%%%

Data reduction was performed using the local software tool \textsc{reduce}, which flags interference, shadowing and hardware errors, applies phase and amplitude calibrations, and Fourier transforms the lag correlator data to synthesize the frequency channels, before outputting to disc in \textit{uv}-\textsc{fits} format.  Flux calibration was performed using short observations of 3C48, 3C286 or 3C147 near the beginning and end of each run.  The assumed flux densities for 3C286 were calculated from Very Large Array total-intensity measurements provided by R. Perley (private communication), and are consistent with the \citet{1987Icar...71..159R} model of Mars transferred onto absolute scale, using results from the \emph{WMAP} satellite.  The assumed flux densities for 3C48 and 3C147 are based on long-term monitoring with the AMI SA using 3C286 for flux calibration (see Table~\ref{tab:Fluxes-of-3C286}).  A correction for each antenna for changing weather and radiometer performance is also applied using a noise-injection system, the `rain gauge' (see \citealt{zwart2008} for more details).

\begin{table}
\centering
\caption{Assumed I~+~Q flux densities of 3C286, 3C48 and 3C147.} \label{tab:Fluxes-of-3C286}
\begin{tabular}{ccccc}\hline
 Channel & $\bar{\nu}$/GHz & $S^{\rm 3C286}$/Jy & $S^{\rm 3C48}$/Jy & $S^{\rm 3C147}$/Jy \phantom{$S^{\rm{X^{X}}}$} \\ \hline
 3 & 13.88 & 3.74 & 1.89 & 2.72 \\
 4 & 14.63 & 3.60 & 1.78 & 2.58 \\
 5 & 15.38 & 3.47 & 1.68 & 2.45 \\
 6 & 16.13 & 3.35 & 1.60 & 2.34 \\
 7 & 16.88 & 3.24 & 1.52 & 2.23 \\
 8 & 17.63 & 3.14 & 1.45 & 2.13 \\ \hline
\end{tabular}
\end{table}

Many of the automatic flagging routines used to excise interference from AMI data rely on the amplitude of the astronomical source being observed remaining constant throughout the observation.  This is not the case for drift-scan data, where sources drift through the primary beam.  It was found that the automated routines (designed in particular to remove interference spikes) were partially flagging out bright sources, so that their flux densities measured from the final maps were systematically lower than expected compared to flux densities from tracked observations.  To overcome this, an iterative scheme was introduced in which the data are reduced as usual, then the visibility data are averaged over channels and baselines and searched for remaining peaks, indicating the presence of a bright source.  The data are then re-reduced, with the time ranges within which bright sources are found being excluded from the interference flagging routines.

In order to process the continuous drift-scans, pointing centres were set up spaced 10\,arcmin apart in RA and each sample was phase-rotated to the closest pointing centre.  The SA integration time is one second, so this results in $600/(15 \cos(\delta))$ samples per pointing centre, which yields approximately 70 samples at $\delta=55^{\circ}$.  The data were then exported in multi-source \textit{uv}-\textsc{fits} format.  The in-house software package \textsc{Fuse} was used to concatenate visibility data from different observations of the same declination strip, weighting each observation according to its noise level; variability of sources has not been considered (see Section.~\ref{flux accuracy}).

\subsection{Mapping}
\begin{figure}
  \begin{center}
  % This file is generated by the MATLAB m-file laprint.m. It can be included
% into LaTeX documents using the packages graphicx, color and psfrag.
% It is accompanied by a postscript file. A sample LaTeX file is:
%    \documentclass{article}\usepackage{graphicx,color,psfrag}
%    \begin{document}\input{beam_sizes}\end{document}
% See http://www.mathworks.de/matlabcentral/fileexchange/loadFile.do?objectId=4638
% for recent versions of laprint.m.
%
% created by:           LaPrint version 3.16 (13.9.2004)
% created on:           20-Aug-2012 20:57:41
% eps bounding box:     15 cm x 11.25 cm
% comment:              
%
\begin{psfrags}%
\psfragscanon%
\Large
%
% text strings:
\psfrag{s01}[t][t]{\color[rgb]{0,0,0}\setlength{\tabcolsep}{0pt}\begin{tabular}{c}Synthesised beam FWHM (arcsec)\end{tabular}}%
\psfrag{s02}[b][b]{\color[rgb]{0,0,0}\setlength{\tabcolsep}{0pt}\begin{tabular}{c}Fraction\end{tabular}}%
\psfrag{s05}[l][l]{\color[rgb]{0,0,0}Minor axis}%
\psfrag{s06}[l][l]{\color[rgb]{0,0,0}Major axis}%
\psfrag{s07}[l][l]{\color[rgb]{0,0,0}Minor axis}%
\psfrag{s09}[][]{\color[rgb]{0,0,0}\setlength{\tabcolsep}{0pt}\begin{tabular}{c} \end{tabular}}%
\psfrag{s10}[][]{\color[rgb]{0,0,0}\setlength{\tabcolsep}{0pt}\begin{tabular}{c} \end{tabular}}%
%
% xticklabels:
\psfrag{x01}[t][t]{0}%
\psfrag{x02}[t][t]{0.1}%
\psfrag{x03}[t][t]{0.2}%
\psfrag{x04}[t][t]{0.3}%
\psfrag{x05}[t][t]{0.4}%
\psfrag{x06}[t][t]{0.5}%
\psfrag{x07}[t][t]{0.6}%
\psfrag{x08}[t][t]{0.7}%
\psfrag{x09}[t][t]{0.8}%
\psfrag{x10}[t][t]{0.9}%
\psfrag{x11}[t][t]{1}%
\psfrag{x12}[t][t]{50}%
\psfrag{x13}[t][t]{100}%
\psfrag{x14}[t][t]{150}%
\psfrag{x15}[t][t]{200}%
\psfrag{x16}[t][t]{250}%
\psfrag{x17}[t][t]{300}%
\psfrag{x18}[t][t]{350}%
\psfrag{x19}[t][t]{400}%
%
% yticklabels:
\psfrag{v01}[r][r]{0}%
\psfrag{v02}[r][r]{0.1}%
\psfrag{v03}[r][r]{0.2}%
\psfrag{v04}[r][r]{0.3}%
\psfrag{v05}[r][r]{0.4}%
\psfrag{v06}[r][r]{0.5}%
\psfrag{v07}[r][r]{0.6}%
\psfrag{v08}[r][r]{0.7}%
\psfrag{v09}[r][r]{0.8}%
\psfrag{v10}[r][r]{0.9}%
\psfrag{v11}[r][r]{1}%
\psfrag{v12}[r][r]{0}%
\psfrag{v13}[r][r]{0.1}%
\psfrag{v14}[r][r]{0.2}%
\psfrag{v15}[r][r]{0.3}%
\psfrag{v16}[r][r]{0.4}%
\psfrag{v17}[r][r]{0.5}%
%
% Figure:
\resizebox{0.9\linewidth}{!}{\includegraphics{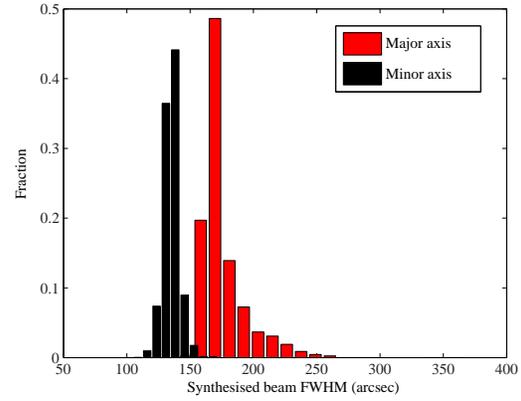}}%
\end{psfrags}%
%
% End beam_sizes.tex

  \caption{The distribution of synthesised beam major and axis FWHMs for the pointing centres which make up the drift-scan maps.}
  \label{Fi:beam_sizes}
  \end{center}
\end{figure}

The pointings were imaged separately in \textsc{Aips}\footnote{\textsc{astronomical image processing system} -- www.aips.nrao.edu/} with $128 \times 128$ pixel maps, where the pixels are $20 \times 20$\,arcsec$^{2}$ in size.  Natural weighting was used to maximise signal-to-noise, and all six frequency channels were imaged using a multi-frequency synthesis; as a consequence of different flagging of the channels, the effective frequency will vary slightly between pointings.  Individual channel maps were not produced.  An automated \textsc{Clean}ing process was used to place a $6\times6$\,pixel \textsc{Clean} box around the brightest pixel if it had a flux density greater than 200\,mJy, and the images were \textsc{Clean}ed to three times the r.m.s. noise on the dirty image; the box was removed, and the \textsc{Clean}ing was continued to the same flux density level.  Each component map was \textsc{Clean}ed using an elliptical Gaussian fitted to the
central region of the dirty beam as the restoring beam.  As a result, the
restoring beam for each component map is slightly different.  The distribution of synthesised beam sizes is shown in Fig.~\ref{Fi:beam_sizes}.

\begin{figure}
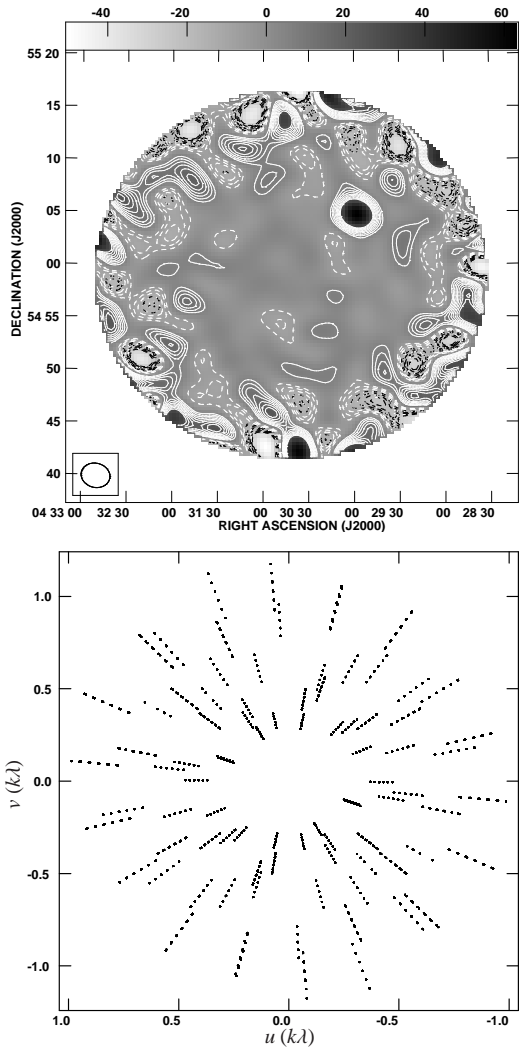

  \begin{center}
  \begin{subfigure}{.9\linewidth}
  \includegraphics[bb=46 143 567 676, clip=, width=0.9\linewidth]{GAL_DRIFT_27.0101.ps}
  \medskip
  \end{subfigure}
  \begin{subfigure}{.9\linewidth}
  \begin{psfrags}%
  \psfragscanon%
  \psfrag{x01}[t][t]{\hspace{25pt}$u\:(k\lambda)$}%
  \psfrag{y01}[b][b]{\hspace{22pt}$v\:(k\lambda)$}%
  \includegraphics[bb=46 125 567 624, clip=, width=0.9\linewidth]{GAL_DRIFT_270101_uv.ps}
  \end{psfrags}
  \end{subfigure}
  \caption{The beam-corrected map (top) and \textit{uv}-coverage (bottom) for a typical drift-scan pointing centre at $\delta \approx 55^{\circ}$.  The 1$\sigma$ map noise at the centre of the map is 2.72\,mJy\,beam$^{-1}$, as estimated by \textsc{Imean}; the noise increases away from the centre (up to a factor of ten).  The map has contours at $\pm2$--10$\sigma$; solid contours are positive and dashed contours are negative.  The synthesised beam is shown in the bottom left-hand corner.  The grey-scale is in units of mJy\,beam$^{-1}$.}
  \label{Fi:pointing_examp}
  \end{center}
\end{figure}

The noise on each component map was estimated using the \textsc{Imean} task over the whole map, which fits a Gaussian centred on
zero to the distribution of pixel values, discarding outliers.  Fig.~\ref{Fi:pointing_examp} shows a typical beam-corrected pointing map at $\delta \approx 55^{\circ}$ with its restoring beam, along with the \textit{uv}-coverage for the pointing centre.

\subsection{Primary-beam correction}
The SA primary beam is usually approximated by a Gaussian fitted to the central lobe of the beam, with a FWHM of 19.6\,arcmin at the central frequency.  However, primary beam correction of drift-scan maps is complicated due to the continuous nature of the scan:  each pointing does not consist solely of data taken towards its centre, but rather of data taken towards a series of points along its RA axis, corresponding to data taken at different times.  The primary beam  was therefore calculated as a weighted average of beams centred at each of the constituent points, i.e. for any pixel in the map

\begin{align}
\textrm{primary beam} &= \frac{\sum_{i=1}^{N} w_i\, \exp \left( -\frac{\Delta_{i}^{2}}{2\sigma^2} \right) }{\sum_{i=1}^{N} w_i} \nonumber\\
 &= \frac{\sum_{i=1}^{N} w_i\, \exp \left( -\frac{(x-x_{i})^{2}+y^{2}}{2\sigma^2} \right)}{\sum_{i=1}^{N} w_i},
\end{align}
where $N$ is the number of samples constituting the pointing, $w_i=1/\sigma_{\mathrm{rms,i}}^2$ is the weight of the $i$'th sample (i.e. the sum of weights for all baselines and all channels contributing to a one-second sample) where $\sigma_{\mathrm{rms,i}}$ is the r.m.s. noise on the sample, $2\sigma\sqrt{(2\ln(2))}$ is the FWHM of the SA primary beam
(19.6\,arcmin), and $\Delta_{i} = \sqrt{(x-x_{i})^{2}+y^2}$ is the separation of the pixel from the pointing centre of the sample, where $(x,y)$ is the pixel location and $(x_{i},0)$ is the pointing centre of the sample along the RA axis.  The pixel value is then divided by the weighted-average beam for that pixel; pixels with a weighted-average beam of $\leq0.1$ are blanked.  This has the effect of elongating the beam along the RA axis to $\approx 37$\,arcmin between the 10\%-power points, compared to the normal SA primary beam RA width (to 10\%) of $\approx 35$\,arcmin.

This beam correction was applied to each of the individual pointing maps produced by \textsc{Aips} and a noise map for each pointing was also produced, which is the inverse of the primary beam scaled by the r.m.s. noise value of the map calculated by \textsc{Imean}.

\subsection{Map combination}
\begin{figure}
  \begin{center}
    \includegraphics[bb=46 144 570 636, clip=, width=0.9\linewidth]{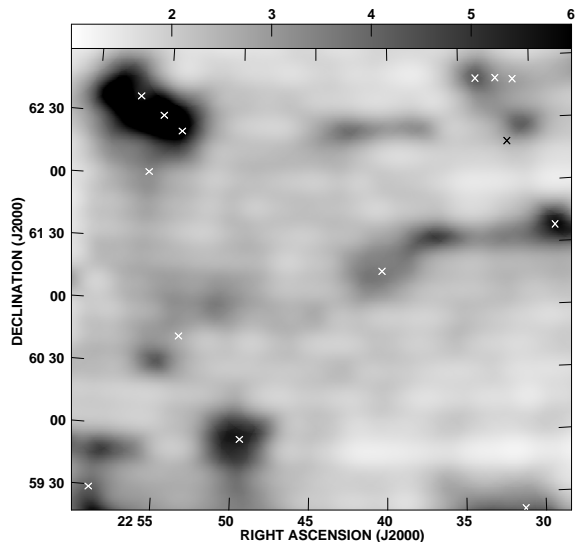}
    \caption{A typical noise map illustrating the variation in noise level across the map.  The grey-scale is in mJy\,beam$^{-1}$ and is truncated at 6\,mJy\,beam$^{-1}$ to show the low-level variation; the highest noise level in the area shown is $\approx10$\,mJy\,beam$^{-1}$ in the north-eastern corner.  Crosses mark the positions of sources with peak flux densities $>50$\,mJy\,beam$^{-1}$, around which it can be seen that the noise level increases.  Away from the bright sources, the noise level is $\lessapprox3$\,mJy\,beam$^{-1}$.}
    \label{Fi:noise_examp}
  \end{center}
\end{figure}

Finally, the individual beam-corrected pointing maps were added together, weighting each pixel by the inverse of its variance calculated from the noise map, into larger continuous maps using the in-house software \textsc{Profile} \citep{2002MNRAS.333..318G}.  Corresponding continuous noise maps for use in source-finding were also produced in the same way from the noise maps for the individual pointing centres; these are found to provide an accurate representation of the noise, except around bright sources as discussed in Section~\ref{Source exclusion}.  Fig.~\ref{Fi:noise_examp} shows an example noise map section illustrating the variation in noise level across a typical map.  All maps were also regridded into Galactic co-ordinates using the \textsc{Aips} task \textsc{regrd}.

%%%%%%%%%%%%%%%%%%%%%%%%%%%%%%%%%%%%%%%%%%%%%%%%%%%%%%%%%%%%%%%%%%%
\section{Source extraction}\label{Source extraction}
%%%%%%%%%%%%%%%%%%%%%%%%%%%%%%%%%%%%%%%%%%%%%%%%%%%%%%%%%%%%%%%%%%%

\citet{2011MNRAS.415.2699A} describes the source extraction methodology used for the 10C survey; a similar process was used for the AMIGPS.  Source finding was carried out over the combined maps to search for peaks greater than 5$\sigma$, where $\sigma$ is the r.m.s. noise value read from the corresponding pixel on the noise map.  A peak position and flux density value is measured by interpolating between the grid points.  An initial estimate of the integrated flux density and source size is also calculated by integrating contiguous pixels down to $2.5\times$ the local thermal noise level, and sources are identified as overlapping if the integration area contains more than one peak $>5\sigma$.  This information is used to fit an elliptical Gaussian to each source in an automated fashion, using the \textsc{Aips} task \textsc{Jmfit}.  Overlapping sources are fitted simultaneously.  As \textsc{Jmfit} can only fit up to four sources simultaneously, in a limited number of cases where more than four sources were identified as overlapping by the source-finding software sources were regrouped manually into groups of four or less.

\subsection{Source size and classification}\label{Source size}
The deconvolved source size is calculated by \textsc{Jmfit} using the synthesised beam size at the pointing with the highest weight at the position of the source.  This size is used to classify the source as point-like or extended to the SA beam, following the method described in \citet{2011MNRAS.415.2699A}, scaled to the SA beam size.  A source is classified as extended if the fitted major axis size $e_{\mathrm{maj}} \geq e_{\mathrm{crit}}$,
where
%%%%%%%%%%%%%%%%%%%%%%%%%%
\begin{eqnarray}
\label{eqn:ecrit}
e_{\mathrm{crit}} =
\left\{
\begin{array}{ll}
3.0 b_{\mathrm{maj}}\,\rho^{-1/2} & \mathrm{if~} 3.0 b_{\mathrm{maj}}\,\rho^{-1/2} > 100.0~\mathrm{arcsec,} \\
100.0~\mathrm{arcsec} & \mathrm{otherwise,}
\end{array}
\right.
\end{eqnarray}
%%%%%%%%%%%%%%%%%%%%%%%%%% 
where $\rho$ is the signal-to-noise ratio and $b_{\mathrm{maj}}$ is the synthesised-beam major-axis size.

If a source is classified as extended, its integrated flux density fitted by \textsc{Jmfit} is considered to best represent its total flux density; otherwise the peak flux density is considered to provide a more accurate measurement.

The error inherent in using the beam from the pointing with the highest weight at the position of the source for source extraction has been investigated by remapping a section of the survey with identical restoring beams for all pointing centres.  The flux densities derived from this map were compared with the catalogue values for sources which lie between pointing centres with different beam shapes and sizes.  For point-like sources, the difference in the flux density is $\lessapprox 1$\% and is considered to be negligible.  For extended sources it is $\lessapprox 5$\%, so a conservative extra 5\% error on the flux density is added in quadrature (see Section~\ref{flux accuracy}).

Although a Gaussian is a reasonable approximation to the shape of many sources, clearly in the Galactic plane there are many sources which are not Gaussian in shape, including complex sources which are treated as multiple overlapping discrete sources.  Integrated flux densities should therefore be used with caution.  The `$\chi^2$' statistic is included in the catalogue as an indication of the goodness of fit, calculated as

\begin{equation}
\chi^2 = \frac{\sum_{i=1}^{N} (S_{i} - \bar{S}_{i})^{2}}{\sigma^{2} (N - 6\times N_{\mathrm{src}})}
\end{equation}
where $N$ is the number of pixels in the fitting area, $S_{i}$ and $\bar{S}_{i}$ are the actual and modelled flux densities of pixel $i$ respectively, $\sigma$ is the estimated thermal noise at the position of the source, and $N_{\mathrm{src}}$ is the number of sources fitted simultaneously, for each of which 6 parameters (central RA, $\delta$, $S_{\mathrm{pk}}$, major and minor axis size and position angle) are fitted.   This should be treated as an indicator, rather than a formal reduced $\chi^2$ since it does not take into account the number of independent pixels in the fitting area, and the value of the noise is uncertain and likely underestimated around bright sources, as described in Section~\ref{Source exclusion}.

For extended sources, users should note that flux densities will be biased low due to the interferometric nature of the survey; the larger spatial scales are under-sampled resulting in flux being `resolved out'.  Any comparison of flux densities of extended objects with other catalogues must take this into account.

%%%%%%%%%%%%%%%%%%%%%%%%%%%%%%%%%%%%%%%%%%%%%%%%%%%%%%%%%%%%%%%%%%%
\subsection{Spurious source exclusion}\label{Source exclusion}
\begin{figure}
  \begin{center}
  % This file is generated by the MATLAB m-file laprint.m. It can be included
% into LaTeX documents using the packages graphicx, color and psfrag.
% It is accompanied by a postscript file. A sample LaTeX file is:
%    \documentclass{article}\usepackage{graphicx,color,psfrag}
%    \begin{document}\input{falloff_laprint}\end{document}
% See http://www.mathworks.de/matlabcentral/fileexchange/loadFile.do?objectId=4638
% for recent versions of laprint.m.
%
% created by:           LaPrint version 3.16 (13.9.2004)
% created on:           07-Aug-2012 15:54:44
% eps bounding box:     15 cm x 11.25 cm
% comment:              
%
\begin{psfrags}%
\psfragscanon%
\Large
%
% text strings:
\psfrag{s07}[t][t]{\color[rgb]{0,0,0}\setlength{\tabcolsep}{0pt}\begin{tabular}{c}Distance from source (arcsec)\end{tabular}}%
\psfrag{s08}[b][b]{\color[rgb]{0,0,0}\setlength{\tabcolsep}{0pt}\begin{tabular}{c}Flux density (Jy\,beam$^{-1}$)\end{tabular}}%
%
% xticklabels:
\psfrag{x01}[t][t]{0}%
\psfrag{x02}[t][t]{0.1}%
\psfrag{x03}[t][t]{0.2}%
\psfrag{x04}[t][t]{0.3}%
\psfrag{x05}[t][t]{0.4}%
\psfrag{x06}[t][t]{0.5}%
\psfrag{x07}[t][t]{0.6}%
\psfrag{x08}[t][t]{0.7}%
\psfrag{x09}[t][t]{0.8}%
\psfrag{x10}[t][t]{0.9}%
\psfrag{x11}[t][t]{1}%
\psfrag{x12}[t][t]{-5000}%
\psfrag{x13}[t][t]{-4000}%
\psfrag{x14}[t][t]{-3000}%
\psfrag{x15}[t][t]{-2000}%
\psfrag{x16}[t][t]{-1000}%
\psfrag{x17}[t][t]{0}%
\psfrag{x18}[t][t]{1000}%
\psfrag{x19}[t][t]{2000}%
\psfrag{x20}[t][t]{3000}%
\psfrag{x21}[t][t]{4000}%
\psfrag{x22}[t][t]{5000}%
%
% yticklabels:
\psfrag{v01}[r][r]{0}%
\psfrag{v02}[r][r]{0.1}%
\psfrag{v03}[r][r]{0.2}%
\psfrag{v04}[r][r]{0.3}%
\psfrag{v05}[r][r]{0.4}%
\psfrag{v06}[r][r]{0.5}%
\psfrag{v07}[r][r]{0.6}%
\psfrag{v08}[r][r]{0.7}%
\psfrag{v09}[r][r]{0.8}%
\psfrag{v10}[r][r]{0.9}%
\psfrag{v11}[r][r]{1}%
\psfrag{v12}[r][r]{-0.05}%
\psfrag{v13}[r][r]{0}%
\psfrag{v14}[r][r]{0.05}%
\psfrag{v15}[r][r]{0.10}%
\psfrag{v16}[r][r]{0.15}%
\psfrag{v17}[r][r]{0.20}%
\psfrag{v18}[r][r]{0.25}%
%
% Figure:
\resizebox{0.9\linewidth}{!}{\includegraphics{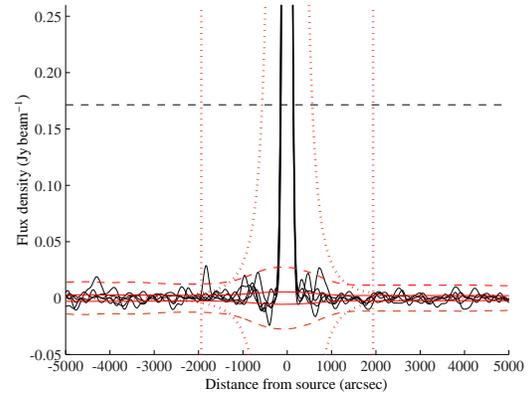}}%
\end{psfrags}%
%
% End falloff_laprint.tex

  \caption{Pixel values (solid black lines) interpolated through map points in lines intersecting the bright, central source in Fig.~\ref{Fi:excl_zones}, and the spurious sources around it; the mean noise- and 5$\sigma$-detection-levels (solid and dashed red lines) from the noise map; the fall-off law and exclusion zone radius for this source (red curved and vertical dotted lines); and the $S_{\mathrm{peak,bright}}/10$ cutoff line (dashed black line).  It can be seen that the noise outside the exclusion zones is well represented by the map noise, however closer to the central source the noise is elevated and the 5$\sigma$ cutoff is not high enough.  The conservative $S_{\mathrm{peak,bright}}/10$ cutoff excludes the spurious detections.}
  \label{Fi:falloff}
\end{center}
\end{figure}

\begin{figure}
  \begin{center}
    \includegraphics[bb=46 112 570 707, clip=, width=0.9\linewidth]{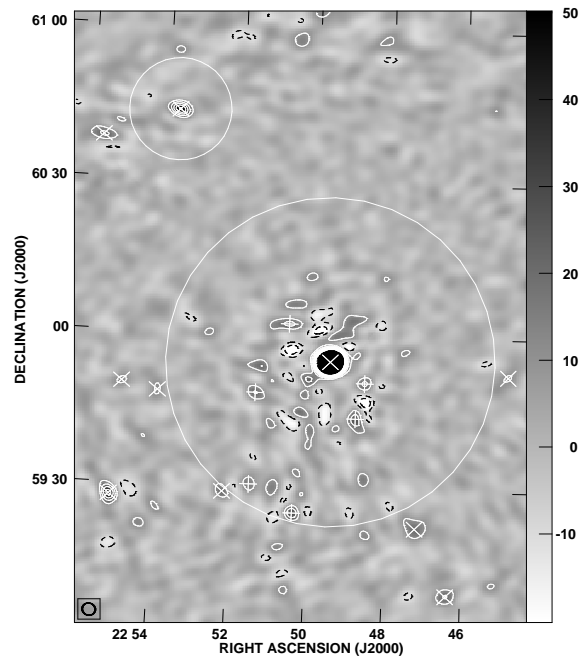}
    \caption{A section of the map illustrating the spurious source exclusion method.  The grey-scale is in mJy\,beam$^{-1}$ and is truncated to show the fainter sources; the flux densities of the brightest and second brightest sources are $\approx$1700\,mJy and 50\,mJy respectively.  The contour levels are between $\pm100$\,mJy\,beam$^{-1}$ in steps of 10\,mJy\,beam$^{-1}$ (it is not possible to use $\sigma$ contours since the noise level varies across the map); solid contours are positive and dashed contours are negative.  Exclusion zones are shown as circles around the bright sources.  Source detections are marked by $\times$, and `sources' detected but excluded by $+$.  The synthesised beam at the position of the brightest source is shown in the bottom left hand corner.}
    \label{Fi:excl_zones}
  \end{center}
\end{figure}

%%%%%%%%%%%%%%%%%%%%%%%%%%%%%%%%%%%%%%%%%%%%%%%%%%%%%%%%%%%%%%%%%%%
An implausibly large number of sources are frequently detected in the vicinity of bright sources -- these are likely spurious and are caused by residual amplitude and phase errors in the data and un\textsc{Clean}ed sidelobes.  In order to prevent these from contaminating the catalogue, `exclusion zones' were applied to sources with peak flux density $>50$\,mJy beam$^{-1}$.  The radii $r_{E}$ of the exclusion zones are determined by the peak flux density $S_{\mathrm{peak,bright}}$ of the source as $r_{E} = 18 \left(S_{\mathrm{peak,bright}} / 300\, \mathrm{mJy}\right)^{1/3}$\,arcmin.  This was chosen empirically to describe the fall-off in the elevated, non-Gaussian noise around bright sources, illustrated in Fig.~\ref{Fi:falloff}.  Within the exclusion zones, only `sources' with peak flux density $S_{\mathrm{peak}} \geq S_{\mathrm{peak,bright}}/10$ are retained.  The factor of ten was conservatively chosen by eye to retain most of the sources which appear to be real, while excluding as many spurious sources as possible.  There may be some real sources which are excluded by this procedure; the implications of this for the completeness of the catalogue will be discussed in Paper~II.  Fig.~\ref{Fi:excl_zones} illustrates the exclusion zones around two bright sources.

%%%%%%%%%%%%%%%%%%%%%%%%%%%%%%%%%%%%%%%%%%%%%%%%%%%%%%%%%%%%%%%%%%%
\section{Calibration accuracy checks}\label{Calibration accuracy checks}
%%%%%%%%%%%%%%%%%%%%%%%%%%%%%%%%%%%%%%%%%%%%%%%%%%%%%%%%%%%%%%%%%%%
As many bright sources found in the Galactic plane are well known and frequently used as phase-calibrator sources by AMI, it is possible to use them to check both the positional and flux calibration accuracy of the drift-scan survey.  A follow-up campaign with the AMI LA has also been conducted on sources identified as having rising spectra with respect to the NVSS \citep{1998AJ....115.1693C} (this will be described in detail in Paper II).  Since the positional accuracy of the LA is much greater than that of the SA (the synthesised beam is $\approx$30\,arcsec), these pointed follow-up observations provide an additional check on the calibration accuracy.

\subsection{Positional accuracy}

\subsubsection{Point-like sources}

The catalogue of source positions (for point-like sources only) derived from the survey maps was matched to the milliarcsecond-accurate positions from the VCS catalogue (\citealt{2002ApJS..141...13B}), resulting in 125 matches with signal-to-noise ratio (SNR) in the drift-scan maps ranging from $\approx8$ to 600.  In addition, the positions derived from follow-up observations of objects that were also point-like to the LA were compared to the drift-scan catalogue positions, resulting in 270 additional matches (not matched to a VCS source) with SNR in the drift-scan maps ranging from $\approx5$ to 400.

The errors $\sigma_{RA}$ and $\sigma_{\delta}$ in RA and $\delta$ for a point source are assumed to be given by 
\begin{subequations}
\label{eqn:pos_err}
\begin{align}
\sigma_{\mathrm{RA}}^2 &= \epsilon_{\mathrm{RA}}^2 + \sigma_{\mathrm{M}}^2 \sin^2(\phi) + \sigma_{\mathrm{m}}^2 \cos^2(\phi) \\
\sigma_{\delta}^2 &= \epsilon_{\delta}^2 + \sigma_{\mathrm{M}}^2 \cos^2(\phi) + \sigma_{\mathrm{m}}^2 \sin^2(\phi),
\end{align}
\end{subequations}
where $\epsilon_{\mathrm{RA}\: or\: \delta}$ are the r.m.s. calibration errors in RA and $\delta$, $\sigma_{\mathrm{M\: or\: m}}$ are the noiselike uncertainties parallel to the synthesised beam major (M) and minor (m) axes, and $\phi$ is the position angle of the beam.  We assume the noiselike uncertainties are given by
\begin{equation}
\label{eqn:pos_err2}
\sigma_{\mathrm{M\:or\: m}} = \frac{\theta_{\mathrm{M\: or\: m}}}{\sqrt{2\ln(2)} \: \mathrm{SNR}},
\end{equation}
where $\theta_{\mathrm{M\: or\: m}}$ are the major and minor FWHM of the synthesised beam.

To test for systematic RA and $\delta$ offsets, the mean offsets between both the AMIGPS and VCS catalogue and AMIGPS and LA positions were calculated separately and as a single group, and by selecting sources with SNR $>$50 and SNR $>$100 in the drift-scan maps.  These are listed in Table~\ref{tab:pos_err}, and are all consistent with zero within $<2.5\sigma$, so we assume no systematic offset in RA or $\delta$.

\begin{table}
\centering
\caption{Mean RA and $\delta$ position offsets for high SNR sources in the drift-scan catalogue (all in arcsec).  Consistency is checked by using the offsets from VCS catalogue and LA positions separately, and combined, and by changing the minimum SNR.} \label{tab:pos_err}

\begin{tabular}{ccccc}\hline
SNR & Offset & Number of & Mean RA & Mean $\delta$ \\
limit & from & sources & offset & offset \\ \hline
\multirow{3}{*}{50} & VCS & 56 & $0.9\pm0.5$ & $-0.4\pm0.3$ \\
 & LA & 18 & $-1.1\pm0.7$ & $-0.9\pm0.7$ \\
 & Combined & 74 & $0.5\pm0.4$ & $-0.5\pm0.3$ \\ \hline
\multirow{3}{*}{100} & VCS & 30 & $0.6\pm0.7$ & $-0.6\pm0.4$ \\
 & LA & 5 & $-0.5\pm1.5$ & $-1.0\pm1.0$ \\
 & Combined & 35 & $0.5\pm0.4$ & $-0.65\pm0.3$ \\
\hline 
\end{tabular}
\end{table}

\begin{figure}
  \begin{center}
    % This file is generated by the MATLAB m-file laprint.m. It can be included
% into LaTeX documents using the packages graphicx, color and psfrag.
% It is accompanied by a postscript file. A sample LaTeX file is:
%    \documentclass{article}\usepackage{graphicx,color,psfrag}
%    \begin{document}\input{all_errs_ea_2.6_ed_1.7_laprint}\end{document}
% See http://www.mathworks.de/matlabcentral/fileexchange/loadFile.do?objectId=4638
% for recent versions of laprint.m.
%
% created by:           LaPrint version 3.16 (13.9.2004)
% created on:           10-Aug-2012 12:51:47
% eps bounding box:     15 cm x 14.3933 cm
% comment:              
%
\begin{psfrags}%
\psfragscanon%
\Large
% text strings:
\psfrag{s01}[t][t]{\color[rgb]{0,0,0}\setlength{\tabcolsep}{0pt}\begin{tabular}{c}RA offset / $\sigma_{RA}$\end{tabular}}%
\psfrag{s02}[b][b]{\color[rgb]{0,0,0}\setlength{\tabcolsep}{0pt}\begin{tabular}{c}$\delta$ offset / $\sigma_{\delta}$\end{tabular}}%
%
% xticklabels:
\psfrag{x01}[t][t]{0}%
\psfrag{x02}[t][t]{0.1}%
\psfrag{x03}[t][t]{0.2}%
\psfrag{x04}[t][t]{0.3}%
\psfrag{x05}[t][t]{0.4}%
\psfrag{x06}[t][t]{0.5}%
\psfrag{x07}[t][t]{0.6}%
\psfrag{x08}[t][t]{0.7}%
\psfrag{x09}[t][t]{0.8}%
\psfrag{x10}[t][t]{0.9}%
\psfrag{x11}[t][t]{1}%
\psfrag{x12}[t][t]{-4}%
\psfrag{x13}[t][t]{-2}%
\psfrag{x14}[t][t]{0}%
\psfrag{x15}[t][t]{2}%
\psfrag{x16}[t][t]{4}%
%
% yticklabels:
\psfrag{v01}[r][r]{0}%
\psfrag{v02}[r][r]{0.1}%
\psfrag{v03}[r][r]{0.2}%
\psfrag{v04}[r][r]{0.3}%
\psfrag{v05}[r][r]{0.4}%
\psfrag{v06}[r][r]{0.5}%
\psfrag{v07}[r][r]{0.6}%
\psfrag{v08}[r][r]{0.7}%
\psfrag{v09}[r][r]{0.8}%
\psfrag{v10}[r][r]{0.9}%
\psfrag{v11}[r][r]{1}%
\psfrag{v12}[r][r]{-4}%
\psfrag{v13}[r][r]{-2}%
\psfrag{v14}[r][r]{0}%
\psfrag{v15}[r][r]{2}%
\psfrag{v16}[r][r]{4}%
%
% Figure:
\resizebox{0.9\linewidth}{!}{\includegraphics{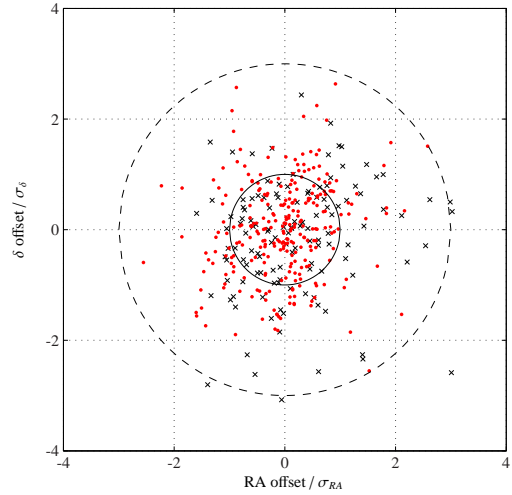}}%
\end{psfrags}%
%
% End all_errs_ea_2.6_ed_1.7_laprint.tex

    \caption{RA and $\delta$ offsets normalised by their calculated errors for all sources which are matched to a VCS source with well-known position (black crosses) or have been followed up with the LA (red dots).  The estimated 1 and 3$\sigma$ error circles are also shown.}
    \label{Fi:offsets}
  \end{center}
\end{figure}

To determine the r.m.s. calibration errors, $\epsilon_{\mathrm{RA}}$ and $\epsilon_{\delta}$ were varied until 99.7\% of the sources with VCS positions had offsets within $3\sigma$ calculated from Equation~\ref{eqn:pos_err}.  This gave $\epsilon_{\mathrm{RA}}=2.6$\,arcsec and  $\epsilon_{\delta}=1.7$\,arcsec.  Fig.~\ref{Fi:offsets} shows the positional offsets for all sources in both datasets, normalised by the calculated error.  They agree well, with 99\% of all offsets lying within the 3$\sigma$ circle (the statistics are only approximately Gaussian).

\subsubsection{Extended sources}
For sources that are extended relative to the SA beam, the positional uncertainty is calculated slightly differently.  The errors in RA and $\delta$ are given by

\begin{subequations}
\label{eqn:pos_err_ext}
\begin{align}
\sigma_{\mathrm{RA}}^2 &= \epsilon_{\mathrm{RA}}^2 + \sigma_{\mathrm{J,RA}}^2 \\
\sigma_{\delta}^2 &= \epsilon_{\delta}^2 + \sigma_{\mathrm{J,\,}\delta}^2,
\end{align}
\end{subequations}
where the $\sigma_{\mathrm{J,RA\:or\:J,\,}\delta}$ terms are the errors estimated by the \textsc{Aips} fitting task \textsc{Jmfit}, which folds in an estimate of the noiselike error as well as the error associated with the fit.

\subsection{Flux calibration accuracy}\label{flux accuracy}

We assume flux calibration errors are given by

\begin{subequations}
\label{eqn:flux_err}
\begin{align}
\sigma_{S_{\mathrm{peak}}}^2 &= (0.05\,S_{\mathrm{peak}})^2 + \sigma^2 \quad &\textrm{for a point-like source} \\
\sigma_{S_{\mathrm{int}}}^2 &= 2\,(0.05\,S_{\mathrm{int}})^2 + \sigma^2 \quad &\textrm{for an extended source},
\end {align}
\end{subequations}
where $S_{\mathrm{peak}}$ is peak flux density and $S_{\mathrm{int}}$ is integrated flux density.  This error estimation comprises a 5\% calibration uncertainty (including rain-gauge correction) and a noise-like error $\sigma$ which for a point-like source is the r.m.s map noise measured from the \textsc{Clean}ed map, and for an extended source is the error estimated by \textsc{Jmfit} which also folds in an estimation of the fitting error.  The error for an extended source also contains an extra 5\% error due to the uncertainty in the beam shape, as discussed in Section~\ref{Source extraction}.  This does not account for the effect of flux loss, as mentioned in Section~\ref{Source size}.

At 16\,GHz, intrinsic source variability is important.  \citet{2009MNRAS.400..995F} find that of 93 extra-galactic sources monitored with the AMI SA for periods between one and 18 months, $\approx$50\% are variable above the flux density calibration uncertainties and 15\% are variable at a level of more than 20\%.  Variability must therefore be considered when attempting to test the flux calibration accuracy.

\subsubsection{NGC~7027}
NGC~7027 is a planetary nebula lying within the drift-scan survey area and for present purposes essentially non-variable (see, e.g. \citealt{2008ApJ...681.1296Z}).  It is also frequently monitored by AMI with tracked observations so an accurate flux density at 16\,GHz can be calculated for comparison.  Using data taken between 2007 and July 2012 with the SA, the 16\,GHz flux of NGC~7027 is 5.4\,Jy.  The drift-scan flux for NGC~7027 is 5.1$\pm$0.3\,Jy, agreeing with the tracked value to within 1.2$\sigma$.

\subsubsection{3C48}
3C48 is one of the primary calibration sources used by AMI and is also known to be non-variable.  Separate drift-scan observations were made of an area around it between Mar and Dec 2010 as an initial test of the drift-scan pipeline. These observations were reduced both in the standard pipeline, which uses the closest primary calibrator observations in time including 3C48, as well as using only 3C286 as a primary calibrator.  3C48 has a flux density of 1.64\,Jy at 16\,GHz; the drift-scan flux density is 1.60$\pm$0.08\,Jy, using 3C48 and 3C286 as primary calibrators, and 1.63$\pm$0.08\,Jy using only 3C286.  Both values are consistent with each other and are within 0.5$\sigma$ of the nominal value.

\subsubsection{Concurrent observations}
\begin{figure}
  \begin{center}
    % This file is generated by the MATLAB m-file laprint.m. It can be included
% into LaTeX documents using the packages graphicx, color and psfrag.
% It is accompanied by a postscript file. A sample LaTeX file is:
%    \documentclass{article}\usepackage{graphicx,color,psfrag}
%    \begin{document}\input{flux_comp_laprint}\end{document}
% See http://www.mathworks.de/matlabcentral/fileexchange/loadFile.do?objectId=4638
% for recent versions of laprint.m.
%
% created by:           LaPrint version 3.16 (13.9.2004)
% created on:           21-Oct-2012 13:33:41
% eps bounding box:     15 cm x 11.25 cm
% comment:              
%
\begin{psfrags}%
\psfragscanon%
\Large
% text strings:
\psfrag{s02}[t][t]{\color[rgb]{0,0,0}\setlength{\tabcolsep}{0pt}\begin{tabular}{c}Mean quasi-simultaneous tracked flux density (mJy)\end{tabular}}%
\psfrag{s03}[b][b]{\color[rgb]{0,0,0}\setlength{\tabcolsep}{0pt}\begin{tabular}{c}(Drift scan)/(Mean tracked) flux density\end{tabular}}%
%
% xticklabels:
\psfrag{x01}[t][t]{0}%
\psfrag{x02}[t][t]{0.1}%
\psfrag{x03}[t][t]{0.2}%
\psfrag{x04}[t][t]{0.3}%
\psfrag{x05}[t][t]{0.4}%
\psfrag{x06}[t][t]{0.5}%
\psfrag{x07}[t][t]{0.6}%
\psfrag{x08}[t][t]{0.7}%
\psfrag{x09}[t][t]{0.8}%
\psfrag{x10}[t][t]{0.9}%
\psfrag{x11}[t][t]{1}%
\psfrag{x12}[t][t]{$10^{3}$}%
\psfrag{x13}[t][t]{$10^{4}$}%
%
% yticklabels:
\psfrag{v01}[r][r]{0}%
\psfrag{v02}[r][r]{0.1}%
\psfrag{v03}[r][r]{0.2}%
\psfrag{v04}[r][r]{0.3}%
\psfrag{v05}[r][r]{0.4}%
\psfrag{v06}[r][r]{0.5}%
\psfrag{v07}[r][r]{0.6}%
\psfrag{v08}[r][r]{0.7}%
\psfrag{v09}[r][r]{0.8}%
\psfrag{v10}[r][r]{0.9}%
\psfrag{v11}[r][r]{1}%
\psfrag{v12}[r][r]{0.7}%
\psfrag{v13}[r][r]{0.8}%
\psfrag{v14}[r][r]{0.9}%
\psfrag{v15}[r][r]{1}%
\psfrag{v16}[r][r]{1.1}%
\psfrag{v17}[r][r]{1.2}%
%
% Figure:
\resizebox{0.9\linewidth}{!}{\includegraphics{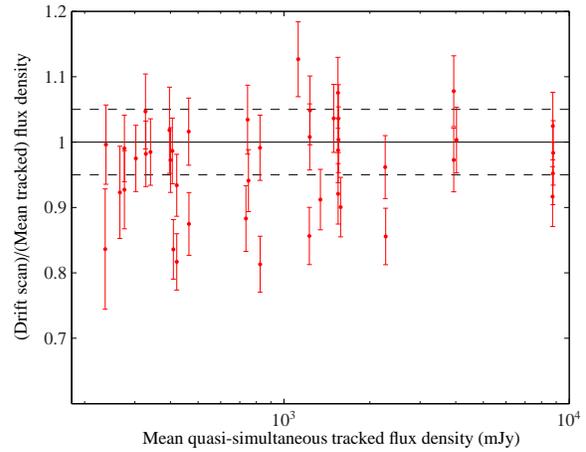}}%
\end{psfrags}%
%
% End flux_comp_laprint.tex

    \caption{Drift-scan flux densities compared to the mean flux from tracked SA archival observations within 10 days.  The black solid and dotted lines show a one-to-one correspondence and $\pm$5\% flux-calibration uncertainty.}
    \label{Fi:flux_comp}
  \end{center}
\end{figure}

Since AMI is continually observing phase calibrators for many of its observations, there is a high probability of quasi-simultaneous tracked observations existing of bright compact sources, mostly extra-galactic, seen in the drift-scan survey.  Extrapolating from Fig.~3 of \citet{2009MNRAS.400..995F} which shows the variability index for extra-galactic sources at 15\,GHz as a function of time, 10 days was chosen as an interval within which source variability should be small.  Since the drift-scan survey also consists of multiple observations at different dates, each observation which contained a potential match within $\pm$10 days was reimaged separately and source-finding was done on the individual declination strips.    Any archival SA tracked observations within $\pm$10 days of drift observations of matching sources were averaged and compared with the individual drift-scan values.  Fig.~\ref{Fi:flux_comp} illustrates the comparison between the peak flux densities of these sources; 93\% of the drift-scan flux densities are within 3$\sigma$ of the mean archival flux.

The three outliers had lower drift-scan flux densities than the mean archival flux density and were found to lie near the edge of the declination strip, where phase errors are expected to have the greatest effect.  In each case, the source appears near the centre of the adjacent strip, which was observed a day later.  When creating the final combined map, the pixels nearer the centre of individual pointings are given greater weight, so the discrepant flux densities will be down-weighted.  The flux densities for these sources derived from raster maps produced from observations close in time agree with the mean archival flux to within 1$\sigma$.

It is common for survey flux densities to be slightly suppressed due to small phase errors shifting the positions of sources which lie away from the pointing centres in the constituent maps (see, e.g. \citealt{2011MNRAS.415.2708A}).  The concurrent observations were tested for this effect, but the median percentage difference ($(S_{\mathrm{mean,tracked}}-S_{\mathrm{drift}})/S_{\mathrm{mean,tracked}}$) was found to be only $\approx2$\%, so the AMIGPS flux densities were not adjusted for this effect.
 
\subsubsection{Non-concurrent observations}
A final check of the flux calibration accuracy can be made by comparing the LA follow-up flux densities to the drift-scan flux densities for sources that are found to be point-like to the LA, although these observations are widely spaced in time (by up to $\approx1.5$ years).  Very little is known about variability statistics in the Galactic plane at cm-wavelength.  However, some idea of the expected number of variable sources can be obtained using results from the 5-GHz Galactic plane variability study by \citep{2010AJ....140..157B}, where $\approx 8$\% of sources detected in the flux density
range from 1 -- 100\,mJy between $b\approx \pm 1.0^{\circ}$ were found to be variable on a time scale of years or shorter when no correction for the inclusion of the extra-galactic source population was applied.

\begin{figure}
  \begin{center}
    % This file is generated by the MATLAB m-file laprint.m. It can be included
% into LaTeX documents using the packages graphicx, color and psfrag.
% It is accompanied by a postscript file. A sample LaTeX file is:
%    \documentclass{article}\usepackage{graphicx,color,psfrag}
%    \begin{document}\input{flux_comp_laprint}\end{document}
% See http://www.mathworks.de/matlabcentral/fileexchange/loadFile.do?objectId=4638
% for recent versions of laprint.m.
%
% created by:           LaPrint version 3.16 (13.9.2004)
% created on:           25-Jul-2012 12:17:49
% eps bounding box:     15 cm x 13.969 cm
% comment:              
%
\begin{psfrags}%
\psfragscanon%
\Large
% text strings:
\psfrag{s01}[t][t]{\color[rgb]{0,0,0}\setlength{\tabcolsep}{0pt}\begin{tabular}{c}LA flux (mJy)\end{tabular}}%
\psfrag{s02}[b][b]{\color[rgb]{0,0,0}\setlength{\tabcolsep}{0pt}\begin{tabular}{c}Drift scan flux (mJy)\end{tabular}}%
%
% xticklabels:
\psfrag{x01}[t][t]{0}%
\psfrag{x02}[t][t]{0.1}%
\psfrag{x03}[t][t]{0.2}%
\psfrag{x04}[t][t]{0.3}%
\psfrag{x05}[t][t]{0.4}%
\psfrag{x06}[t][t]{0.5}%
\psfrag{x07}[t][t]{0.6}%
\psfrag{x08}[t][t]{0.7}%
\psfrag{x09}[t][t]{0.8}%
\psfrag{x10}[t][t]{0.9}%
\psfrag{x11}[t][t]{1}%
\psfrag{x12}[t][t]{10}%
\psfrag{x13}[t][t]{100}%
\psfrag{x14}[t][t]{1000}%
\psfrag{x15}[t][t]{10000}%
%
% yticklabels:
\psfrag{v01}[r][r]{0}%
\psfrag{v02}[r][r]{0.1}%
\psfrag{v03}[r][r]{0.2}%
\psfrag{v04}[r][r]{0.3}%
\psfrag{v05}[r][r]{0.4}%
\psfrag{v06}[r][r]{0.5}%
\psfrag{v07}[r][r]{0.6}%
\psfrag{v08}[r][r]{0.7}%
\psfrag{v09}[r][r]{0.8}%
\psfrag{v10}[r][r]{0.9}%
\psfrag{v11}[r][r]{1}%
\psfrag{v12}[r][r]{10}%
\psfrag{v13}[r][r]{100}%
\psfrag{v14}[r][r]{1000}%
\psfrag{v15}[r][r]{10000}%
%
% Figure:
\resizebox{0.9\linewidth}{!}{\includegraphics[bb=23 25 390 380, clip=]{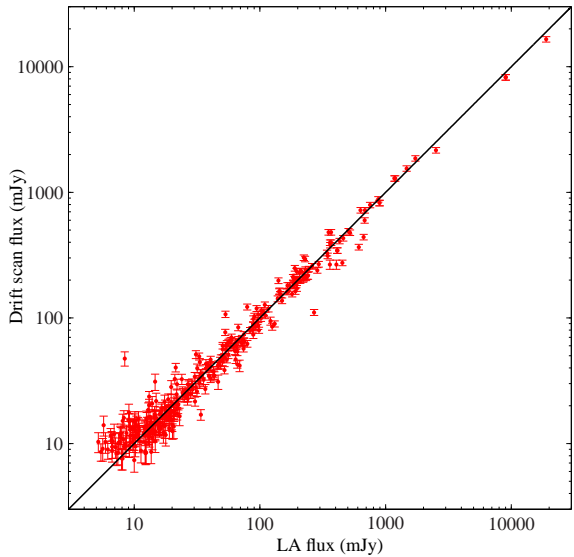}}%
\end{psfrags}%
%
% End flux_comp_laprint.tex

    \caption{Drift-scan flux densities compared to the LA follow-up flux.  The black solid line shows a one-to-one correspondence.}
    \label{Fi:flux_comp_la}
  \end{center}
\end{figure}

Fig.~\ref{Fi:flux_comp_la} shows the comparison between the LA and drift-scan survey peak flux densities.  87\% are within 3$\sigma$, taking into account the LA errors which are generally smaller than the drift-scan errors and are not plotted for clarity.  The remaining 13\% seems consistent with the 8\% of sources predicted to be variable, given that no correction for differences in frequency, flux density range, Galactic latitude distribution or bias due to selecting for rising spectrum sources has been attempted.  The apparent bias towards higher drift-scan flux densities at the lower end of the flux density scale is likely an Eddington bias caused by low-SNR sources selected from the AMIGPS map being more likely to occur on positive noise bumps.

%%%%%%%%%%%%%%%%%%%%%%%%%%%%%%%%%%%%%%%%%%%%%%%%%%%%%%%%%%%%%%%%%%%
\section{Data products}\label{Data products}
%%%%%%%%%%%%%%%%%%%%%%%%%%%%%%%%%%%%%%%%%%%%%%%%%%%%%%%%%%%%%%%%%%%

%%%%%%%%%%%%%%%%%%%%%%%%%%%%%%%%%%%%%%%%%%%%%%%%%%%%%%%%%%%%%%%%%%%
\subsection{Raster maps}\label{Raster maps}
%%%%%%%%%%%%%%%%%%%%%%%%%%%%%%%%%%%%%%%%%%%%%%%%%%%%%%%%%%%%%%%%%%%

\begin{figure*}
 \begin{center}
  % This file is generated by the MATLAB m-file laprint.m. It can be included
% into LaTeX documents using the packages graphicx, color and psfrag.
% It is accompanied by a postscript file. A sample LaTeX file is:
%    \documentclass{article}\usepackage{graphicx,color,psfrag}
%    \begin{document}\input{raster_positions}\end{document}
% See http://www.mathworks.de/matlabcentral/fileexchange/loadFile.do?objectId=4638
% for recent versions of laprint.m.
%
% created by:           LaPrint version 3.16 (13.9.2004)
% created on:           19-Aug-2012 17:18:02
% eps bounding box:     15 cm x 6.7895 cm
% comment:              
%
\begin{psfrags}%
\psfragscanon%
%
% text strings:
\psfrag{s01}[t][t]{\color[rgb]{0,0,0}\setlength{\tabcolsep}{0pt}\begin{tabular}{c}Galactic longitude (degrees)\end{tabular}}%
\psfrag{s02}[b][b]{\color[rgb]{0,0,0}\setlength{\tabcolsep}{0pt}\begin{tabular}{c}Galactic latitude (degrees)\end{tabular}}%
%
% xticklabels:
\psfrag{x01}[t][t]{0}%
\psfrag{x02}[t][t]{0.1}%
\psfrag{x03}[t][t]{0.2}%
\psfrag{x04}[t][t]{0.3}%
\psfrag{x05}[t][t]{0.4}%
\psfrag{x06}[t][t]{0.5}%
\psfrag{x07}[t][t]{0.6}%
\psfrag{x08}[t][t]{0.7}%
\psfrag{x09}[t][t]{0.8}%
\psfrag{x10}[t][t]{0.9}%
\psfrag{x11}[t][t]{1}%
\psfrag{x12}[t][t]{60}%
\psfrag{x13}[t][t]{80}%
\psfrag{x14}[t][t]{100}%
\psfrag{x15}[t][t]{120}%
\psfrag{x16}[t][t]{140}%
\psfrag{x17}[t][t]{160}%
\psfrag{x18}[t][t]{180}%
%
% yticklabels:
\psfrag{v01}[r][r]{0}%
\psfrag{v02}[r][r]{0.1}%
\psfrag{v03}[r][r]{0.2}%
\psfrag{v04}[r][r]{0.3}%
\psfrag{v05}[r][r]{0.4}%
\psfrag{v06}[r][r]{0.5}%
\psfrag{v07}[r][r]{0.6}%
\psfrag{v08}[r][r]{0.7}%
\psfrag{v09}[r][r]{0.8}%
\psfrag{v10}[r][r]{0.9}%
\psfrag{v11}[r][r]{1}%
\psfrag{v12}[r][r]{-5}%
\psfrag{v13}[r][r]{0}%
\psfrag{v14}[r][r]{5}%
%
% Figure:
\resizebox{0.9\linewidth}{!}{\includegraphics{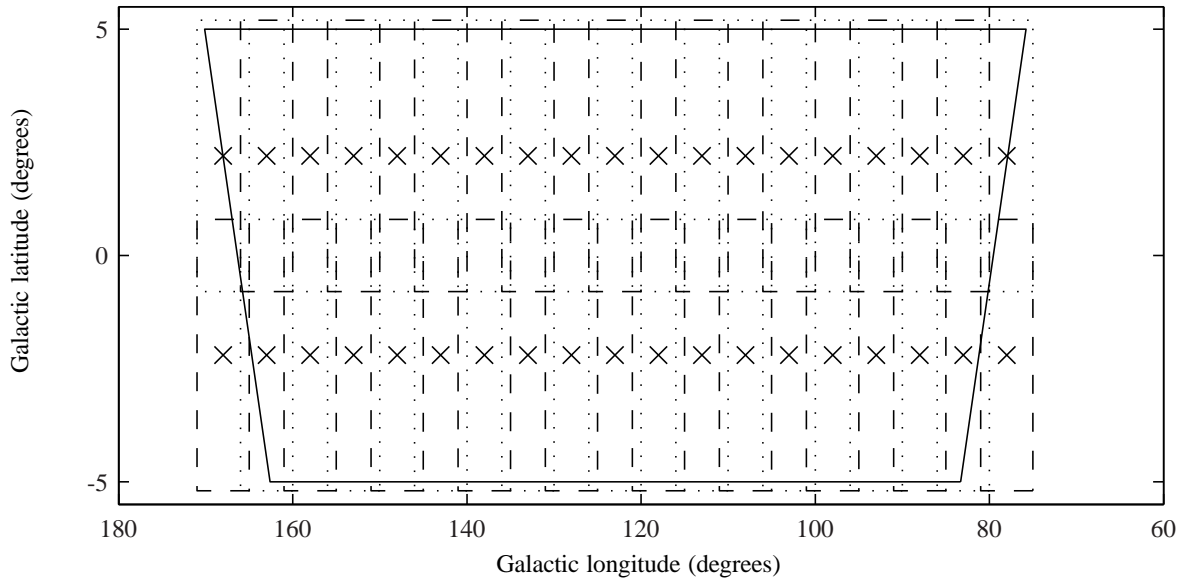}}%
\end{psfrags}%
%
% End raster_positions.tex

  \caption{The positions of the raster maps in Galactic coordinates.  The solid black line marks the extent of the data, the dotted and dashed lines show the boundaries of the raster maps and the crosses mark the centres of the maps.}
  \label{fig:raster_positions}
 \end{center}
\end{figure*}

The field is divided into 38 square maps of side 6$^{\circ}$, and are given names constructed from the Galactic coordinates of their centres, eg G78.0$-$2.2.  These are shown in Fig.~\ref{fig:raster_positions}.  The centres are spaced by 5$^{\circ}$ in longitude, and 4.4$^{\circ}$ in latitude, and start at $\ell = 78.0^{\circ}$, $b = -2.2^{\circ}$.

These raster maps are available from http://www.mrao.cam.ac.uk/surveys/AMIGPS/ or http://vizier.u-strasbg.fr/viz-bin/VizieR, along with:
\begin{itemize}
\item{noise maps containing the estimated thermal noise level at each pixel;}
\item{noise maps adjusted for the exclusion zones around the bright sources.  For a given pixel, the value is $\max(\mathrm{thermal\: noise}, S_{\mathrm{peak,bright}}/50)$, i.e. the (flux-detection limit)/5 for the catalogue;}
\item{a \textsc{fits} data-cube giving the synthesised beam major and minor axis FWHM and position angle appropriate to each pixel (i.e. the synthesised beam belonging to the pointing with the highest weight at that pixel).}
\end{itemize}

\begin{figure}
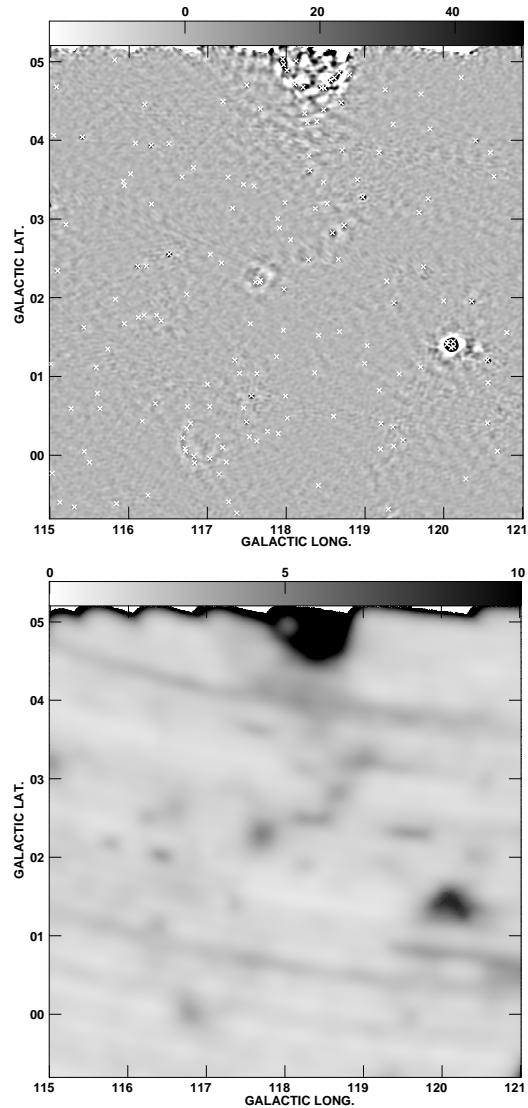

  \begin{center}
  \begin{subfigure}{.9\linewidth}
  \includegraphics[bb=48 115 567 663, clip=, width=0.9\linewidth]{AMI.ps}
  \medskip
  \end{subfigure}
  \begin{subfigure}{.9\linewidth}
  \includegraphics[bb=48 115 569 663, clip=, width=0.9\linewidth]{AMI_noise.ps}
  \medskip
  \end{subfigure}
  \caption{An example AMIGPS raster map (top) and noise map (bottom) centred at $\ell = 118.0^{\circ}$, $b = 2.2^{\circ}$.  Source detections are marked with $\times$; the grey-scales of the map and noise map are in mJy\,beam$^{-1}$ and are truncated to show the fainter features.  Fig.~\ref{Fi:CGPS} shows the CGPS 1.4\,GHz total intensity map of the same region.}
  \label{Fi:examp_map}
  \end{center}
\end{figure}

Fig.~\ref{Fi:examp_map} shows an example 6\,deg$^{2}$ map, with annotations marking the sources detected within it.  Also shown for comparison is a CGPS total intensity 1.4\,GHz map showing the same region.  It can be seen that many sources detected by CGPS are also detected by the AMIGPS; however some larger-scale features such as the supernova remnant G116.5+1.1 are resolved out.  As noted in Section~\ref{Source size}, any comparison of AMIGPS flux densities with other catalogues must take into account the spatial scales probed by the instruments; see Fig.~\ref{Fi:pointing_examp} for the typical $uv$-coverage of an AMIGPS pointing centre.

\begin{figure}
  \begin{center}
  \includegraphics[bb=48 115 567 663, clip=, width=0.9\linewidth]{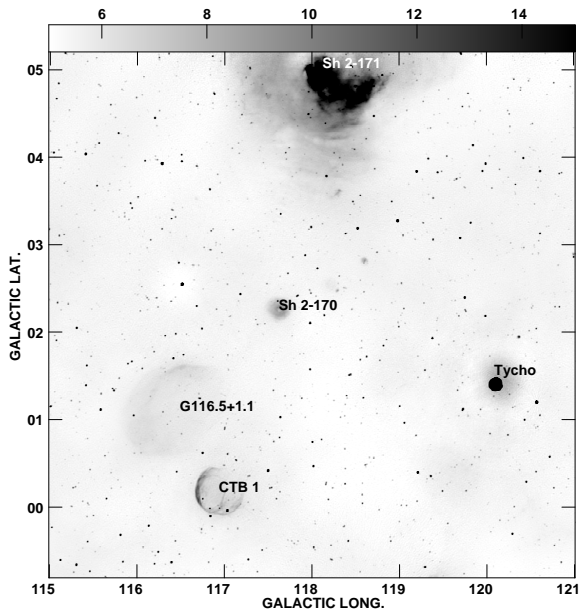}
  \caption{The CGPS 1.4\,GHz total intensity map for the region corresponding to the AMIGPS map shown in Fig.~\ref{Fi:examp_map}.  The grey-scale is in K and is truncated at 15\,K to show the fainter features.  Some well-known supernova remnants (SNR) and \textsc{Hii} regions visible in the map are labelled (\citealt{2009BASI...37...45G}, \citealt{1959ApJS....4..257S}).  It can be seen that the AMIGPS sees many features common to the CGPS, however the larger-scale features such as the SNR G116.5+1.1 are resolved out.}
  \label{Fi:CGPS}
  \end{center}
\end{figure}
%%%%%%%%%%%%%%%%%%%%%%%%%%%%%%%%%%%%%%%%%%%%%%%%%%%%%%%%%%%%%%%%%%%
\subsection{Source catalogue}\label{Source catalogue}
%%%%%%%%%%%%%%%%%%%%%%%%%%%%%%%%%%%%%%%%%%%%%%%%%%%%%%%%%%%%%%%%%%%

A sample of the catalogue containing the first ten sources detected in Fig.~\ref{Fi:examp_map} is shown in Table~4.  The complete source list, which contains 3503 entries, is available as an electronic supplement, from http://vizier.u-strasbg.fr/viz-bin/VizieR or from http://www.mrao.cam.ac.uk/surveys/AMIGPS/.  For each source, the catalogue contains:

\begin{itemize}
\item{A source name, constructed from the J2000 RA and $\delta$ coordinates of the source.}
\item{The peak RA, $\delta$, flux density and associated errors (these are the appropriate quantities to use for point-like sources).}
\item{The fitted centroid RA and $\delta$, integrated flux density and associated errors (these are the appropriate quantities to use for extended sources).}
\item{The critical source size as defined in Eqn.~\ref{eqn:ecrit} and the deconvolved source major and minor axis sizes and position angle.  A deconvolved size of 0.0 indicates that the source was not found to be wider than the synthesised beam in the major or minor axis direction.}
\item{The $\chi^2$ value for the fit.}
\item{The source classification (point-like or extended).}
\end{itemize}

%\begin{table*}
%\vbox to220mm{\vfil Landscape table to go here.}
%\caption{} 
%\vfil{}
%\label{tab:examp_map}
%\end{table*}

\begin{figure*}
\includegraphics{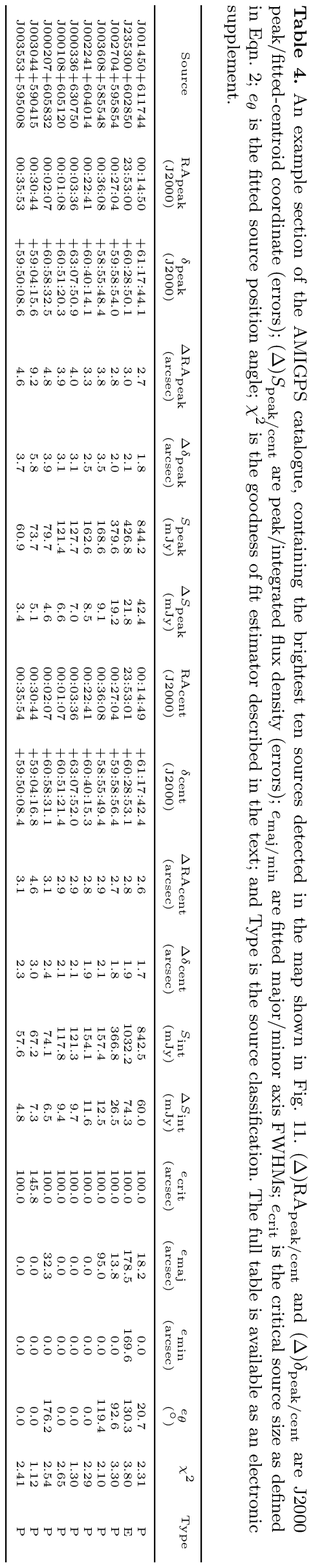}
\end{figure*}

%%%%%%%%%%%%%%%%%%%%%%%%%%%%%%%%%%%%%%%%%%%%%%%%%%%%%%%%%%%%%%%%%%%
\section{Conclusions}
%%%%%%%%%%%%%%%%%%%%%%%%%%%%%%%%%%%%%%%%%%%%%%%%%%%%%%%%%%%%%%%%%%%

The Galactic plane between $b\approx\pm5^{\circ}$ has been surveyed using the interferometric AMI SA at $\approx$16\,GHz, to a noise level of $\approx3$\,mJy\,beam$^{-1}$ at $\approx3$\,arcmin resolution.  This is the most sensitive and highest resolution Galactic plane survey at cm-wave frequencies above 1.4\,GHz.

\begin{enumerate}
\item{868\,deg$^{2}$ of the Galactic plane have been surveyed and a catalogue of 3503 sources produced.  This is the first data release of the AMIGPS.}
\item{As part of creating the AMIGPS, we have developed an automated pipeline to produce maps from data taken in drift-scan mode, accounting for the presence of bright sources.}
\item{The source extraction techniques developed for the 10C survey have been applied to maps at different resolution and regions of the sky with many extended sources present.}
\item{In testing the flux calibration of the survey by comparing source flux densities derived from the AMIGPS to tracked observations of both extra-galactic and Galactic sources taken with the AMI SA and AMI LA, we find that the AMIGPS flux calibration is accurate to within 5\%.}
\item{The r.m.s. positional accuracy of the survey, assessed by comparing positions derived from the AMIGPS with well-known source positions from the VLBA calibrator survey and with AMI LA follow-up positions, is 2.6\,arcsec in RA and 1.7\,arcsec in $\delta$.}
\end{enumerate}

In a following paper the first results from the survey will be presented, and in a future data release the survey will be extended to $\delta\geq20^{\circ}$.

%%%%%%%%%%%%%%%%%%%%%%%%%%%%%%%%%%%%%%%%%%%%%%%%%%%%%%%%%%%%%%%%%%%
\section*{Acknowledgments}\label{acknowledgements}
%%%%%%%%%%%%%%%%%%%%%%%%%%%%%%%%%%%%%%%%%%%%%%%%%%%%%%%%%%%%%%%%%%%

We thank the staff of the Mullard Radio Astronomy Observatory for their invaluable assistance in the commissioning and operation of AMI, which is supported by Cambridge University and the STFC.  We also thank the referee, Robert Watson for helpful comments.  This work has made use of the distributed computation grid of
the University of Cambridge (CAMGRID).  This research has made use of NASA's Astrophysics Data System Bibliographic Services and the facilities of the Canadian Astronomy Data Centre operated by the National Research Council of Canada with the support of the Canadian Space Agency. The research presented in this paper has used data from the Canadian Galactic Plane Survey a Canadian project with international partners supported by the Natural Sciences and Engineering Research Council. YCP acknowledges the support of a Rutherford Foundation/CCT/Cavendish Laboratory studentship.  MPS and CR acknowledge the support of STFC studentships.

%%%%%%%%%%%%%%%%%%%%%%%%%%%%%%%%%%%%%%%%%%%%%%%%%%%%%%%%%%%%%%%%%%%
\setlength{\labelwidth}{0pt}

\label{lastpage}
\end{document}